\documentclass[preprint,showpacs,preprintnumbers,amsmath,amssymb]{revtex4}

\usepackage{graphicx}

\begin{document}

\title{Polarization Saturation in Strained Ferroelectrics}
\author{Yanpeng Yao}
\author{Huaxiang Fu}
\affiliation{Department of Physics, University of Arkansas,
Fayetteville, AR 72701, USA}
\date{\today}

\begin{abstract}

Using density-functional calculations we study the structure and
polarization response of tetragonal PbTiO$_3$, BaTiO$_3$ and
SrTiO$_3$ in a strain regime that is previously overlooked.
Different from common expectations, we find that the polarizations
in all three substances saturate at large strains, demonstrating a
universal phenomenon. The saturation is shown to originate from
an unusual and strong electron-ion correlation that leads to
cancellation between electronic and ionic polarizations. Our results shed
new insight on the polarization properties, and reveal the existence
of a fundamental limit to the strain-induced polarization
enhancement.


\end {abstract}
\maketitle

Response of electrical polarization to external strains in
infinite solids is a subject of interest from both fundamental and
technological points of view.\cite{Dawber} Fundamentally, the
process is governed by electrons, ions and their complex
interactions, forming an important class of collective phenomena.
Seeking direct physics and/or mechanism underlying the
responses is of key relevance. Previously, Ederer and Spaldin
showed that the strain dependence of polarization in different
substances (BaTiO$_3$, BiFeO$_3$, LiNbO$_3$, and PbTiO$_3$) could
be understood by the piezoelectric and elastic constants of the
{\it unstrained} materials.\cite{Ederer} Technologically, inplane
compressive strains were able to dramatically raise the critical
transition temperature in BaTiO$_3$ (Ref.\onlinecite{Choi}) and to
turn incipient SrTiO$_3$ into strongly ferroelectric \cite{Haeni}.
These important studies focus on small strains near equilibrium.

Nowadays, advance in experimental fabrication of various
high-quality ferroelectric(FE) nanostructures\cite{Scott} (such
as thin films, nanowires, and dots) opens an opportunity for
studying polarization response in a new regime with
medium and/or large strains that have not been previously
investigated \cite{Ederer,Choi,Haeni}. Unlike bulks, FE wires or dots
allow exceedingly large external strains without formation of
dislocation, since stress can be effectively relaxed due to
finite lateral size.\cite{Zou} For instance, large compressive
strains can be realized in FE nanorods\cite{Urban} by compressing them
along the lateral directions. Even for thin films, strains of more
than 3\% were shown possible in LaAlO$_3$/SrTiO$_3$.\cite{Reyren}
The new strain regime is interesting and may
reveal many new physics in the sense that the charge density is
{\it heavily} deformed by large strains, making electrons behave more
collectively in some spatial region, but less in another. As a
result, answers to questions (1) what determine FE off-center
displacements in the new strain regime,
and (2) how piezoelectric response, as a collective
phenomenon, adjusts itself to large strains, remain unknown and are
important. Furthermore, based on the widely accepted conclusion that
polarization is to be enhanced as compressive strain
increases\cite{Ederer,Choi,Haeni},
scientists have been long wondering what is the upper limit that
polarization is able to reach, provided that exceeding strain
is possible in laboratory.\cite{Dkhil} This is an intriguing
question since it defines a profound limit on our widespread
effort of seeking enhanced polarization by strain.\cite{Choi,Haeni}

Here by means of first-principle calculations we investigate the
structure and polarization responses of FE perovskites in the new
strain regime. Three paradigm materials
[namely, PbTiO$_3$ (PT), BaTiO$_3$ (BT), and SrTiO$_3$ (ST)]
are studied, aimed to obtain conclusions that are
generally applicable. We find that polarizations in considered
FEs show a universal behavior, by exhibiting three distinct stages
in their strain-induced responses (each stage producing dissimilar
piezoelectric coefficients). Two out of these three stages are new
and previously unknown. We further reveal that when the
compressive inplane strain is high, the polarization will no longer
increase, defying the common belief that polarization must increase
with increasing compressive strain. In
fact, our calculations predict the existence of a critical
inplane strain, above which the polarization starts to saturate.
Our results suggest that there is a fundamental constraint on
polarization enhancement by mechanical strain.

Three considered materials represent FEs of different magnitudes
of polarization: PbTiO$_3$ is strongly ferroelectric, BaTiO$_3$ is
ferroelectric but with a weaker polarization, and SrTiO$_3$ is not
FE under zero strain. We study the structural phase of
tetragonal symmetry ($|{\bf a}_1|$=$|{\bf a}_2|$=$a$, $|{\bf a}_3|$=$c$),
in which materials can have polarizations along the [001]
direction. Biaxial in-plane strain (in Voigt notation) is
defined as $\eta _1$=$\eta _2$=$(a-a_0)/a_0$, where $a_0$ is the unstrained
in-plane lattice constant.

Polarization in infinite periodic solids can be
expressed in lattice-vector coordinate as ${\bf P}=\frac{e}{\Omega
} (\chi _1{\bf a}_1 + \chi _2{\bf a}_2 + \chi _3{\bf a}_3)$, where
$\Omega $ is the volume of a unit cell and $\chi _i$ describes the
polarization along the direction of lattice vector ${\bf
a}_i$. Displacement of ions and deformation of electron density
both contribute to {\bf P}. Computing of ionic contribution
(labeled as {\bf P}$_{\rm ion}$) is straightforward using point
charges, while electronic contribution (denoted as {\bf P}$_{\rm el}$)
is determined using the modern theory of
polarization\cite{King-Smith,Resta}, as implemented in our
mixed-basis method. More specifically, {\bf P}$_{\rm el}$ is
computed as the geometrical Berry phase of valence electron states
as ${\bf P_{\rm el}}=i\frac{2e}{(2\pi )^3}\int d{\bf k}_1
\nabla _{{\bf k}_2} \phi ({\bf k}_1,{\bf k}_2)|_{{\bf k}_1={\bf
k}_2}$, where $\phi ({\bf k}_1,{\bf k}_2)={\rm ln}\, {\rm det}
|<u_{m{\bf k}_1}|u_{n{\bf k}_2}>|$ is the phase of the determinant
formed by occupied Bloch wave functions $u_{n{\bf k}}$. When
inplane strains are imposed, two factors give rise to the change
in polarization: (1) the variation of $\chi _i$, and (2) the
cell-structure modifications of $\Omega $ and ${\bf a}_i$.
The first factor is critical and is
manifested\cite{DV1,Gotthard} in the {\it proper} piezoelectric
coefficients as $e^{\rm
pro}_{\alpha \beta \gamma }=\frac{e}{\Omega }\sum _i \frac{d\chi
_i}{d\epsilon _{\beta \gamma }}a_{i\alpha }$, where $\epsilon
_{\beta \gamma }$ is the strain in Cartesian notation. The
second factor contributes to the improper piezoelectric
coefficient which is trivially related to the proper one by the
magnitude of polarization as $e^{\rm impro} _{\alpha \beta \gamma
}=e^{\rm pro}_{\alpha \beta \gamma } +\delta _{\alpha
\beta}P_{\gamma }-\delta _{\beta \gamma }P_{\alpha }$. One thus
notes that $\chi _i$ is a critical quantity that is
technologically relevant in terms of obtaining high piezoelectric
efficiency. This study will be concerned with the strain
dependence of $\chi _i$.

Technically, we use density functional theory within the local
density approximation (LDA)\cite{Hohenberg} to determine
optimized cell shape and atomic positions by minimizing the total
energy. Calculations are performed using pseudopotential
method with mixed-basis set.\cite{Fu_mix} The Troullier-Martins
type of pseudopotentials are used\cite{Troullier}; atomic
configurations for generating pseudopotentials,
pseudo/all-electron matching radii, and accuracy checking were
described elsewhere.\cite{Vpseudo}  For each in-plane strain, the
out-of-plane $c$ lattice constant and atomic positions are fully
relaxed. Our LDA-calculated inplane lattice constants of unstrained
bulks are $a$=3.88{\AA} for PT, $a$=3.95{\AA} for BT, and
$a$=3.86{\AA} for ST, which all agree well with other existing
calculations.

\begin{figure}
\centering
\includegraphics[width=15cm]{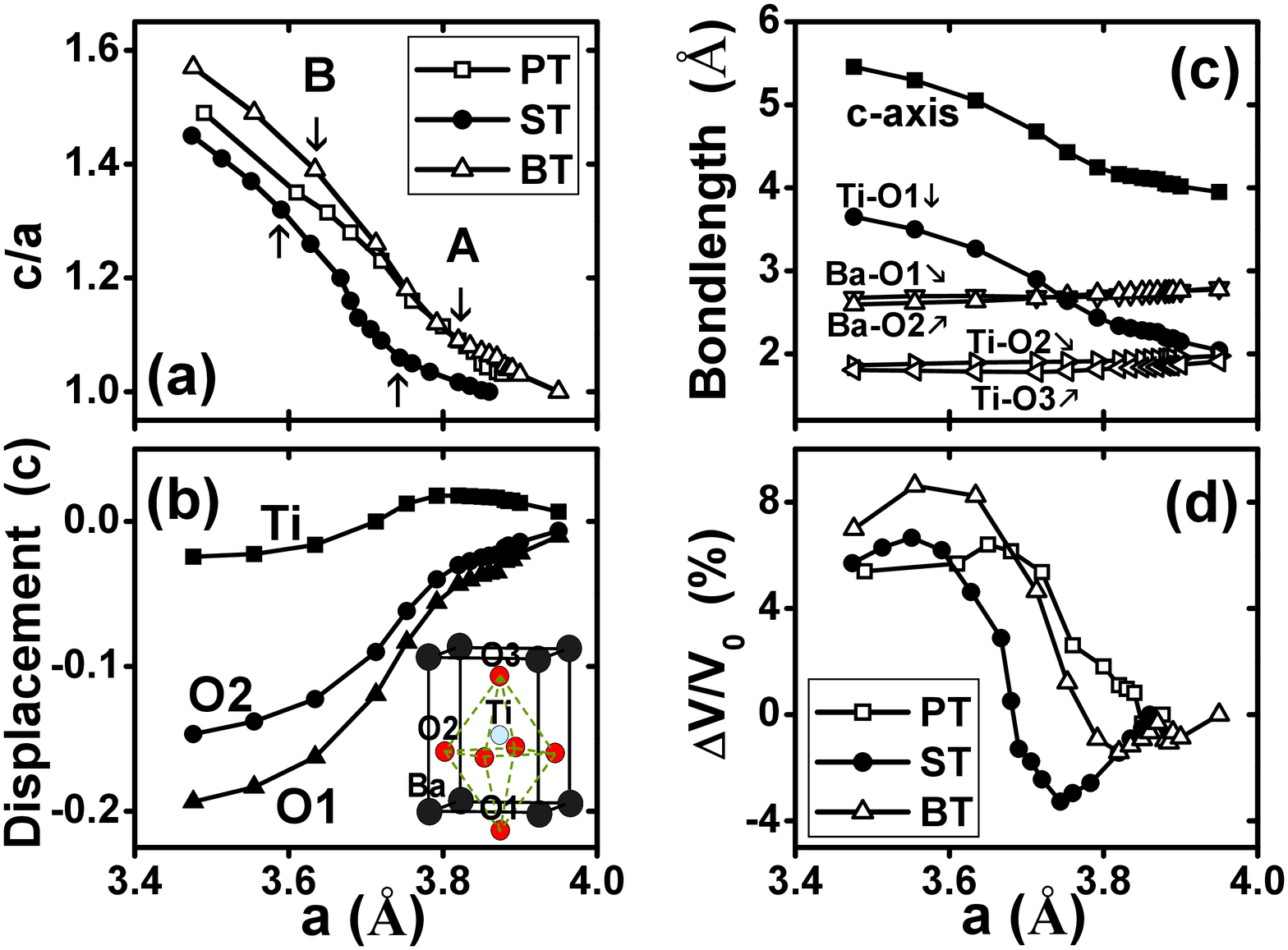}
\caption{Structural properties as a function of inplane lattice
constant: (a) the $c/a$ ratios (where points A and B are marked only
for BT and ST, for the clarity of display), (b) atomic displacements in
BaTiO$_3$, (c) bond lengths in BaTiO$_3$ (where the $c$-axis length is
also plotted for comparison), (d) the change of cell
volume $\Delta V/V_0$ (where $V_0$ is the equilibrium volume).
Labels of atoms in tetragonal perovskite are given in the inset of (b).} \label{Fstr}
\end{figure}

{\it Structural response:} Fig.\ref{Fstr}(a) depicts the $c/a$ ratios
of the three considered materials under compressive strains. Let
us examine the BaTiO$_3$ curve first. For this curve, (1) our
calculations confirm the (expected) linear relation between $c/a$
and $a$ when inplane strain is small; (2) However, as the inplane
$a$ length is decreased to 3.82{\AA} [point A in
Fig.\ref{Fstr}(a)], BaTiO$_3$ apparently enters a different state,
as witnessed by the fact that the $c/a$ rises more sharply afterwards; (3)
Interestingly, when $a$ reaches point B, the $c/a$ becomes linear
again. These results combine to make the $c/a$ in the region between
A and B appear to be notably enhanced. This enhanced $c/a$ regime
in FEs has not been emphasized before (since previous studies
deal with small strains)---and provides an explanation for the (mysterious)
enlarged electromechanical response recently observed in PZT films under
very large electric fields\cite{Grigoriev}, because these fields drive the films
into the new strain range between A and B.

Contrast of three materials in Fig.\ref{Fstr}(a) tells us that the
above conclusions are applicable also for PbTiO$_3$ and SrTiO$_3$.
But the $c/a$ value and the location of point A reveal interesting
differences. First, we predict that for a fixed $a$ constant,
BaTiO$_3$ exhibits a {\it larger} c/a than PbTiO$_3$, unlike the
zero-strain case where PT has a much higher $c/a$ ratio of $\sim
$1.04 (in LDA). Meanwhile, the $c/a$
in SrTiO$_3$ is considerably lower than in BT or PT, which
could be attributed to the small size of Sr. When strained, the
big Ba and Pb atoms repel other atoms more forcefully, leading to
large $c$ lengths. This size difference is also reflected in the
location of the critical A point (at which the $c/a$ changes
behavior). The $a$ value of point A is remarkably similar for PT
and BT, but not for ST.

Fig.\ref{Fstr}(b) shows atomic off-center displacements (in unit of $c$) in
BaTiO$_3$ under different strains, obtained by placing the Ba atom at the
cell origin and then determining the displacements of Ti, O1 and O2
atoms with respect to their high-symmetry locations in tetragonal
structure. Labels of individual atoms are given in the
inset of Fig.\ref{Fstr}(b). Atomic displacements in
Fig.\ref{Fstr}(b) explain why point A plays an important role in
determining the structural response. When $a>a_A$ (lower strains),
the O displacements are small. But, as $a$ is below $a_A$, the
oxygen off-center shifts increase dramatically with a markedly
different slope. Furthermore, the Ti displacement is positive (out
of phase with the O displacement) when $a>a_A$, and then becomes
negative (in phase with the O displacement) for $a<a_A$.

Given the complexity in atomic displacements as well as in
the $c/a$ ratio, one would desire to find a
mechanism (if any) that may determine how, and in what magnitudes,
atoms are displaced. For this purpose, we examine the bond lengths
in strained BaTiO$_3$, shown in Fig.\ref{Fstr}(c). Two
observations are ready: (1) The Ti-O1 distance correlates well
with the $c$ length, (2) Despite that $a$ is significantly
compressed [by $\sim $0.5{\AA} in Fig.\ref{Fstr}(a)], the Ti-O2
and Ti-O3 distances remarkably maintain to be nearly constant (so
are the Ba-O1 and Ba-O2 bondlengths). Our calculations thus reveal the
existence of a simple (and preferred) rule that can quantitatively
explain the structural response {\it over a wide range} of inplane
strains, that is, FEs respond to the external strains in a manner
that maintains the lengths of short Ti-O and Ba-O bonds. This rule
is found true also for PT and ST, and may be easily verified in
x-ray experiments.

Nevertheless, FE perovskites do not maintain the bond lengths in a
trivial way. In Fig.\ref{Fstr}(d) we describe the strain dependence of
the optimal cell volume. Though cell volume is often assumed to be constant under
biaxial strains, obviously this is not the case for FEs. Instead,
Fig.\ref{Fstr}(d) displays features that are very
suggestive in terms of understanding the structure and
polarization responses. Under small strains, the cell volume first
decreases. This decrease is most pronounced in SrTiO$_3$. Then
approximately at $a_A$, the volume of the material increases
quickly, and later at $a_B$, it starts to decrease again. One thus
sees that between $a_A$ and $a_B$ the enhanced $c/a$ in
Fig.\ref{Fstr}(a)---and the enlarged O displacements in
Fig.\ref{Fstr}(b)---occur by {\it the expansion of cell volume}.

\begin{figure}
\centering
\includegraphics[width=15cm]{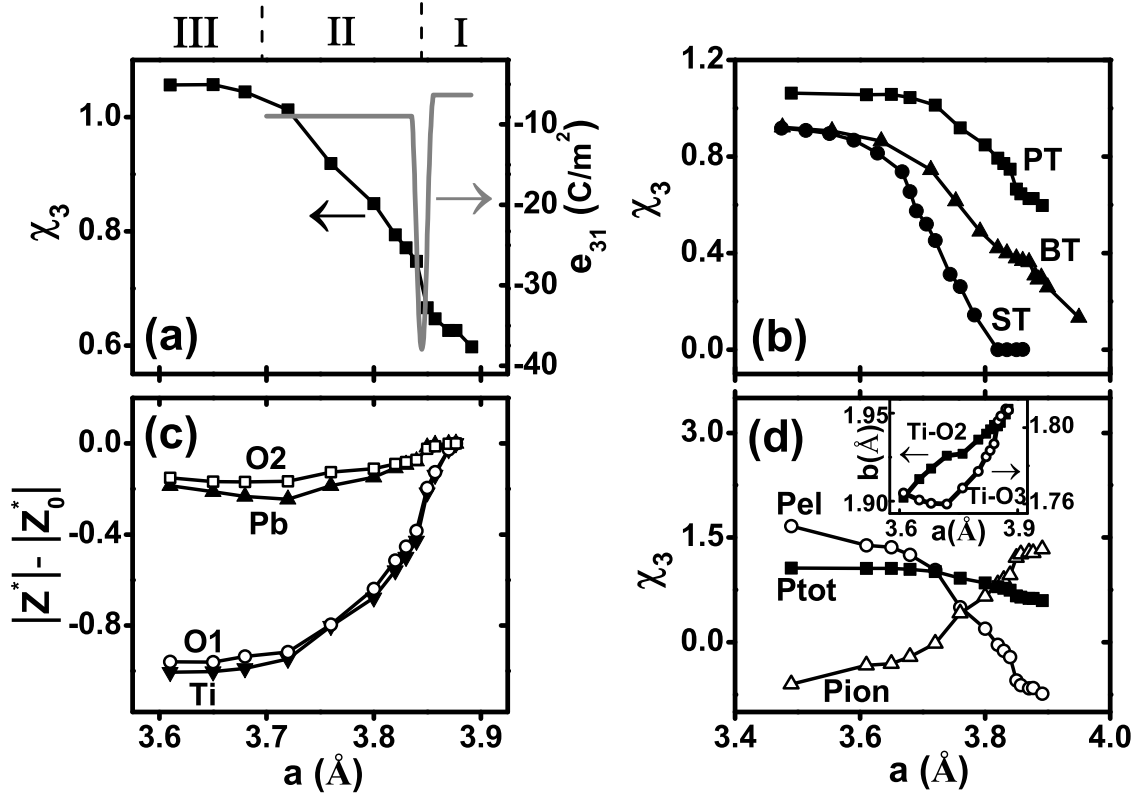}
\caption{Dependencies of the following properties as a function of
the inplane lattice constant: (a) the $\chi _3$ component of
polarization (using the left vertical axis) and piezoelectric
$e^{\rm pro}_{31}$ coefficient (using the right vertical axis) in
PbTiO$_3$, (b) the $\chi _3$ polarizations in BaTiO$_3$ and
SrTiO$_3$, compared with that in PbTiO$_3$, (c) effective charges
$\mid Z^{*}\mid -\mid Z^{*}_{0}\mid$ in PbTiO$_3$,
and (d) electronic (P$_{\rm el}$) and ionic (P$_{\rm ion}$) contributions to the
total (P$_{\rm tot}$) polarization in PbTiO$_3$. For each calculated point in (d), the
A-site atom is always placed at the cell origin so that comparison of
P$_{\rm el}$ at different lattice constants are meaningful.
Inset of (d) shows the fine structure
of Ti-O2 and Ti-O3 bond lengths (b) versus the inplane ($a$) constant.
} \label{Fpol}
\end{figure}

{\it Polarization response:} Fig.\ref{Fpol}(a) depicts the $\chi
_3$ polarization and the proper $e^{\rm pro}_{31}$ piezoelectric
coefficient in PbTiO$_3$ under varied compressive strain. Caution is
made to ensure that we follow continuously the same polarization
as the $\chi _3$ crosses different branches, since polarization in
periodic solids is a multi-valued quantity. At zero strain,
PbTiO$_3$ shows a theoretical $\chi _3$ of 0.63 that corresponds to a
polarization of 66.7$\mu $C/cm$^{2}$, consistent with the experimental
result\cite{Lines} of $\sim $70$\mu $C/cm$^{2}$. Upon
strain, the most notable result in Fig.\ref{Fpol}(a) is that
PbTiO$_3$ displays three stages (labeled as I, II, and III) in
its polarization response.

At stage I (when $a$ is above 3.84{\AA}), the $\chi _3$
polarization increases linearly, from which the piezoelectric
$e^{\rm pro}_{31}$ coefficient is deduced as 637 $\mu $C/cm$^2$.
At stage II, the $\chi _3\sim a$ relation remains linear, but shows
a different slope from that of stage I. We numerically find that the
$e^{\rm pro}_{31}$ for stage II is 898 $\mu $C/cm$^2$.
Interestingly, when the inplane strain continuously transitions from
stage I to II, we notice an abrupt rise in the polarization. This
leads to the occurrence of a huge piezoelectric $e^{\rm pro}_{31}$
coefficient of 3900 $\mu $C/cm$^2$ near $a=3.84${\AA}. 

When the inplane $a$ constant is further decreased below 3.70{\AA}
(which corresponds to a moderate $4.5 \%$ strain), another unusual
phenomenon occurs in Fig.\ref{Fpol}(a). That is, the increase in
polarization dramatically slows down, and the $\chi _3$ finally
{\it saturates} at a value of 1.05. Further increase of inplane
strain no longer affects the $\chi _3$ polarization. This
saturation is surprising, since it is generally accepted that
FE polarization always increases with increasing compressive
strains. The saturated polarization also demonstrates
the existence of a maximal limit up to which
polarization can be enhanced by means of epitaxial strains.

We have also performed polarization calculations for BaTiO$_3$ and
SrTiO$_3$; results are shown in Fig.\ref{Fpol}(b).  For incipient SrTiO$_3$,
we find a threshold strain at $\eta _1$=1\%, below which the
polarization remains null. A similar magnitude of threshold strain
was reported previously.\cite{Antons} Results for BaTiO$_3$ and
SrTiO$_3$ in Fig.\ref{Fpol}(b) reveal that (1) the existence of
a three-stage polarization response and (2) the saturation of
polarization apply to all three substances, hence demonstrating a
general phenomenon. Also interestingly, we find that $\chi _3$
saturates at similar values in BaTiO$_3$ and in SrTiO$_3$, but
differently in PbTiO$_3$.

The saturation of $\chi _3$ can not be naively related to the
saturations of effective charges and/or atomic displacements.
First, we see in Fig.\ref{Fstr}(b) that the O displacements do not
saturate, in fact. To illustrate if effective charge saturates, we
compute the effective $Z^*_{33}$ charge of each atom in PbTiO$_3$
under strains, by finite difference
$Z^*=\frac{\Omega}{e}\frac{\Delta {\bf P}}{\Delta {\bf r}} $. Under
zero strain, the effective charges (denoted as $Z^*_0$) are 3.65
(Pb), 5.76 (Ti), -4.90 (O1), and -2.28 (O2). With strain, we are
interested in the change of the effective charges, namely
$\mid Z^{*}\mid -\mid Z^{*}_{0}\mid$, given in Fig.\ref{Fpol}(c).
Note that although the polarization increases once we turn on
compressive strain, the effective charge nevertheless {\it
decreases} for all atoms. In other words, the $\mid Z^*\mid$ has
its maximum at the equilibrium structure. It is also interesting
to recognize that the $\mid Z^{*}\mid -\mid Z^{*}_{0}\mid$ in
Fig.\ref{Fpol}(c) separate into two groups: one group consists
of Ti and O1 atoms, while the other is formed by Pb and O2 atoms.
The strain dependence of the effective charge is very similar within
the same group, implying the occurrence of a chemical correlation.
Meanwhile, we note that the Z$^*$ do not saturate---and for instance,
the $Z^*$ of Pb (also of O2) slightly increases when the $a$
length is below 3.7{\AA}. Since $\delta \chi _3 = \sum _{i} \{ \delta
Z^*_i\Delta r_{i,z} +Z^*_i\delta (\Delta r_{i,z}) \}$ (where $i$ is
the index of atom), the saturation in $\chi _3$ thus results from
a balanced cancellation between strain-induced variations of $Z^*$ and
atomic displacements.

\begin{figure}
\centering
\includegraphics[width=15cm]{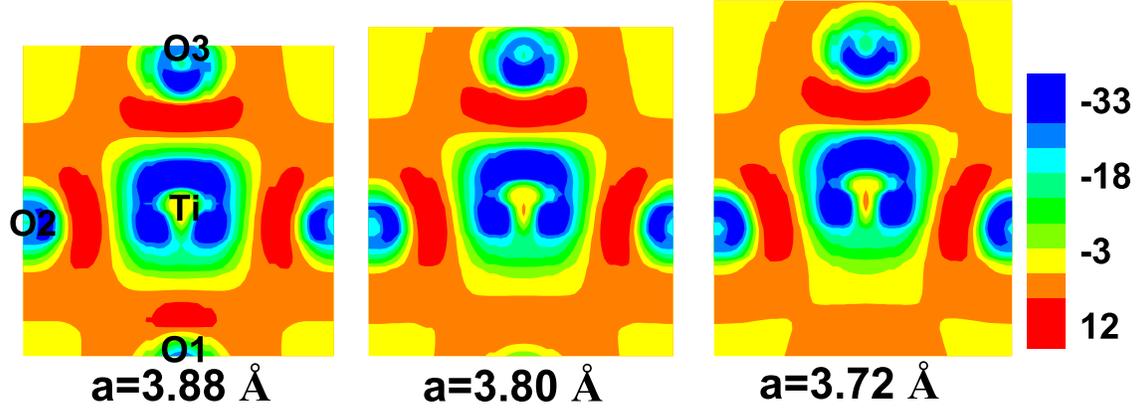}
\caption{(Color online) Charge-density difference $\Delta \rho  =\rho -\rho _0$,
where $\rho_0$ is the sum of electron density of free atoms,
in PbTiO$_3$ at three different inplane lattice constants
of a=3.88, 3.80, and 3.72{\AA}. The plotted density is
on the plane that cuts through the Ti, O1, O2, and O3 atoms.} \label{Fchg}

\end{figure}

We now provide microscopic insight into why polarization undergoes a
(puzzling) change from stage I to stage II, and why $\chi _3$
saturates at high strains, by examining electron charge
distributions (Fig.\ref{Fchg}). The electron distribution in
equilibrium PbTiO$_3$ is as expected: charge transfer leads to
electron deficiency near atomic cores (blue regions), while more charge
accumulates in the interstitial area between atoms (red regions). As PT is strained
to $a$=3.80{\AA}, an important change occurs, namely the
overlapping charge largely disappears between Ti and O1 atoms.
This reveals that the (considerably) weakened Ti-O1 bond is the
reason responsible for the transition from stage I to stage II. It
also explains why the $c/a$ is enhanced in the second stage.

On the other hand, when PbTiO$_3$ is further strained to
$a$=3.72{\AA}, electron charge near the O3 atomic core is {\it heavily}
deformed, as a result of two factors that (1) more electrons are
transferred into the interstitial between Ti and O3, and (2) the
strong Coulomb repulsion from these interstitial charges distort
the electrons near the O3 nuclei. In this circumstance, electrons
between Ti and O3 behave more like strongly correlated particles.
Further, since the electrons near the O3 nuclei couple directly
with this ion, ionic displacement and electron deformation are thus
expected to be correlated as well. This expectation is indeed
confirmed by the electronic (P$_{\rm el}$) and ionic (P$_{\rm ion}$)
contributions to the total (P$_{\rm tot}$) polarization [Fig.\ref{Fpol}(d)].
When $a$ decreases from $a_0$ to 3.72{\AA}, P$_{\rm el}$ rises faster
than P$_{\rm ion}$ declines, giving rise to an overall enhancement of
the total polarization. However, when $a$ is below 3.72{\AA}, the
P$_{\rm el}$ and P$_{\rm ion}$ contributions show equal slopes but with
opposite signs (due to the above-described electron-ion correlation),
thus leading to the polarization saturation
in stage III. Another interesting indication of
the strong electron-ion correlation is manifested by the {\it
fine} structure on how the Ti-O3 bondlength depends on strain [see
the inset of Fig.\ref{Fpol}(d)]. We find that when $\chi _3$
starts to saturate near $a$=3.72{\AA}, the Ti-O3 bondlength
surprisingly begins to {\it increase}. This demonstrates the
existence of a minimum length (B$_{\rm min}$)
by which the Ti-O3 short bond can be compressed. Our predicted
B$_{\rm min}$ is 1.76{\AA} (PT), 1.78{\AA} (BT), and 1.79{\AA} (ST).
Reaching this minimum length implies the onset of polarization saturation.

In summary, density functional calculations on strained FEs reveal
two unusual properties of polarization: (1) the occurrence of
stage II with distinct behaviors, and (2) the
saturation of polarization upon higher strains. We further predict
that the continuous transition from stage I to II may produce
marked piezoelectric responses. Microscopic insights responsible
for these behaviors are discussed. The enhanced piezo-response and
c/a ratio in stage II is associated with the weakening of the
Ti-O1 bonds. On the other hand, the electron-electron and
electron/ion correlations lead to the polarization saturation. Our
calculations reveal the existence of a profound limit up to which
polarization can be enhanced by strain. The predicted saturation
value of polarizations are larger than 125 $\mu $C/cm$^2$ for three
substances, suggesting that currently there is still room for polarization
enhancement. This work was supported by the Office of Naval Research. We thank
H. Krakauer and D. Vanderbilt for very useful discussions.

{\it Note added:} After having completed the manuscript, we were brought
to a recent study\cite{HN_Lee} where the polarizations in highly strained
Pb(Zr$_{0.2}$Ti$_{0.8}$)O$_3$ films (with $c/a\approx$1.09) is surprisingly
similar to that in relaxed PZT films (with $c/a\approx$1.05). This,
in spirit, adds further evidence supporting the validity of our results on polarization
saturation. Nevertheless, our key conclusions that (1) polarization saturates
in all materials, thus being a universal principle; (2) the saturation
origins from strong electron-ion correlation; (3) there is an
unusual volume expansion and an enhanced $c/a$ structural response
between points A and B, have not been addressed previously\cite{HN_Lee}.

\end{document}